\documentclass[aps,prd,nofootinbib,twocolumn,
preprintnumbers]{revtex4}

\usepackage{amssymb}
\usepackage{amsmath}
\usepackage{hyperref}
\usepackage{slashed}
\usepackage{color}
\usepackage{graphicx}
\usepackage{bm}
\usepackage[T1]{fontenc}

\makeatletter
\def\simgt{\mathrel{\lower2.5pt\vbox{\lineskip=0pt\baselineskip=0pt
           \hbox{$>$}\hbox{$\sim$}}}}
\def\simlt{\mathrel{\lower2.5pt\vbox{\lineskip=0pt\baselineskip=0pt
           \hbox{$<$}\hbox{$\sim$}}}}
\makeatother

\newcommand{\be}{\begin{equation}}
\newcommand{\ee}{\end{equation}}
\newcommand{\bea}{\begin{eqnarray}}
\newcommand{\eea}{\end{eqnarray}}
\newcommand{\Eq}[1]{Eq.~(\ref{#1})}

\newcommand{\Fig}[1]{Fig.~\ref{#1}}

\newcommand{\mPl}{m_{\rm Pl}}

\begin{document}

\preprint{CALT 68-2924}
\title{Baryogenesis with Higher Dimension Operators}

\author{Clifford Cheung and Koji Ishiwata}
\affiliation{California Institute of Technology, Pasadena, CA 91125}

\begin{abstract}

  We propose a simple model of baryogenesis comprised of the standard
  model coupled to a singlet $X$ via higher dimension operators ${\cal
    O}$.  In the early universe, $X$ is thermalized by ${\cal O}$
  mediated scattering processes before it decouples relativistically
  and evolves into a sizable fraction of the total energy density.
  Eventually, $X$ decays via ${\cal O}$ in an out of equilibrium,
  baryon number and CP violating process that releases entropy and
  achieves baryogenesis for a broad range of parameters.  The decay
  can also produce a primordial abundance of dark matter.  Because $X$
  may be as light as a TeV, viable regions of parameter space lie
  within reach of experimental probes of $n$-$\bar{n}$ oscillation,
  flavor physics, and proton decay.

\end{abstract}

\maketitle

\begin{center}
\textbf{\textsc{I. Introduction}} 
\end{center}

The standard model (SM) cannot explain the observed matter-antimatter
asymmetry of the universe, and so new physics is required.  In this
letter we propose a simple scenario for baryogenesis consisting of the
SM plus an inert multiplet of states $X$.  These states interact
weakly with the SM through baryon number and CP violating higher
dimension operators ${\cal O}$ set by the scale $\Lambda$.

\begin{figure}[t]
  \begin{center}
    \includegraphics[scale=0.3]{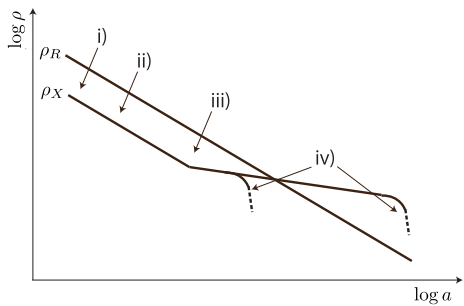}
  \end{center}
  \caption{The four stages of baryogenesis, shown in terms of the
    evolution of the energy density in SM radiation and $X$ as a
    function of scale factor.  The decay of $X$ may occur before or
    after $X$ grows to dominate the total energy density.}
  \label{fig:model}
\end{figure}

\begin{figure}[t]
  \begin{center}
    \includegraphics[scale=0.4]{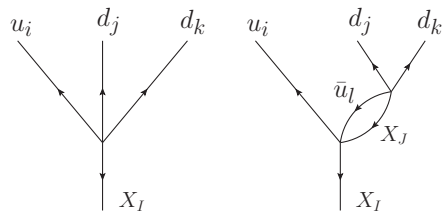}
  \end{center}
  \caption{Tree and one-loop diagrams for $X_I\rightarrow u_i
    d_j d_k$ which interfere to produce a primordial baryon asymmetry. }
  \label{fig:XIudd}
\end{figure}

The process of baryogenesis occurs in the four stages depicted in
\Fig{fig:model}.  In the beginning,
\begin{itemize}
\item[\it i)] 
  $X$ is thermalized with the SM plasma.
\end{itemize}
This condition is possible provided $T_R$, the reheating temperature,
is greater than $m_X$, the mass of $X$.  Hence, thermalization occurs
automatically via scattering in the SM plasma mediated by ${\cal O}$
or the ultraviolet dynamics which generates ${\cal O}$.  Once the
universe cools sufficiently, ${\cal O}$ mediated scattering goes out
of equilibrium and
\begin{itemize}
\item[\it ii)] 
  $X$ decouples relativistically from the SM plasma.
\end{itemize}
Once $X$ leaves equilibrium, it redshifts like radiation until
temperatures drop below $m_X$, at which point $X$ becomes
non-relativistic.  Once $X$ begins to redshift like matter,
\begin{itemize}
\item[\it iii)]
  $X$ evolves into a large fraction of the total energy.
\end{itemize}
During this period the energy density in $X$ is greater than that of
any given relativistic species, and may even come to dominate the
total energy density, sending the universe into a matter dominated
phase.  The epoch of $X$ domination terminates when
\begin{itemize}
\item[\it iv)] $X$ decays, yielding a primordial baryon asymmetry.
\end{itemize}
Crucially, these out of equilibrium decays of $X$ occur via the very
same baryon number and CP violating higher dimension operators ${\cal
  O}$ that initially thermalize $X$ in the early universe.
Interference between tree and one-loop decay amplitudes generate a
baryon asymmetry in the final state, as depicted in \Fig{fig:XIudd}
for an explicit model.  In certain models, $X$ decays can also
generate a primordial abundance of dark matter (DM).

Let us highlight the key features of this baryogenesis scenario.
First, since this mechanism allows for low scale baryogenesis, the
operators ${\cal O}$ may be indirectly probed through $n$-$\bar n$
oscillations, flavor violation, and proton decay.  Second, this setup
is quite minimal, in that only a handful of new particles $X$ are
required and the very same operators ${\cal O}$ that produce $X$
initially also mediate its decay.  As we will see later, a subset of
the $X$ particles can even be DM.  Third, this
setup exploits a cosmological ``fixed point'' arising because $X$ is
{typically} thermalized for a very broad range of reheating
temperatures.
 
As is well-known, the production and thermalization of inert particles
at reheating is a ubiquitous difficulty in theories beyond the SM.
This issue arises in the cosmology of gravitinos
\cite{Moroi:1993mb,Khlopov:1984pf,Kawasaki:1994af,Kawasaki:2008qe,Bolz:2000fu,Pradler:2006qh,Cheung:2011nn,Rychkov:2007uq},
axinos
\cite{Brandenburg:2004du,Covi:2001nw,Strumia:2010aa,Cheung:2011mg},
photini \cite{Ibarra:2008kn,Arvanitaki:2009hb}, and goldstini
\cite{Cheung:2010mc,Cheung:2010qf}.  Transforming this peril into a
blessing is an old idea, {\it e.g.}~in models linking gravitino or
axino domination to baryogenesis in R-parity violating supersymmetry
\cite{Cline:1990bw,Mollerach:1991mu}.  However, we argue that this
mechanism applies much more broadly and is a natural byproduct of
additional singlet states coupled to the SM via baryon number and CP
violating higher dimension operators---the out of equilibrium
condition arises from relativistic decoupling and decays of $X$.
Alternatively, the out of equilibrium condition for $X$ can be
achieved through heavy particle decays \cite{Fukugita:1986hr,
  Lazarides:1991wu,Asaka:1999yd,Jeong:2013axf,Davoudiasl:2010am},
first order phase transitions
\cite{Kuzmin:1985mm,Anderson:1991zb,Cohen:1993nk,Cheung:2012im}, or
rolling scalars \cite{Affleck:1984fy,Dine:1995uk,Dine:2003ax}.
Mechanisms involving higher dimension operators have also been
discussed in more specific contexts \cite{Babu:2006xc,Chung:2001tp}.

In Sec.~II we present a simple example model that illustrates the
salient features of our setup.  We then discuss constraints from
experimental limits in Sec.~III and conclude in Sec.~IV.

\begin{center}
\textbf{\textsc{II. The Model}} 
\end{center}

In this section we present a simple theory illustrating our mechanism
for baryogenesis.  Consider the SM augmented by a multiplet of gauge
singlet Majorana fermions $X_I$ with mass $m_I$.  The interaction
Lagrangian for $X_I$ is
\begin{eqnarray}
{\cal L} =
\frac{\kappa_{IJij}}{\Lambda^2} (X_I u_i)(\bar{X}_J \bar{u}_j)
+\frac{\lambda_{Iijk}}{\Lambda^2}
 (X_I u_i)(d_j d_k) + {\rm c.c.},
\label{eq:exmodel}
\end{eqnarray}
where $i,j,k$ label right-handed quark flavors. Lorentz indices are
contracted implicitly among terms in parentheses, while color indices
are contracted implicitly in the unique way.  It is straightforward to
include other Lorentz and flavor structures into the Lagrangian, but
such terms will not qualitatively alter the mechanics of the model.

In terms of symmetries, baryon number is violated because $X_I$ are
Majorana, while CP is violated because the couplings $\kappa_{IJij}$
and $\lambda_{Iijk}$ are complex.  Note that there is an exact, unbroken
$\mathbb{Z}_2$ subgroup of baryon number under which $X_I$ and the
quarks are all odd.

Let us now describe the cosmological history of this model.  To begin,
we assume that the SM is reheated to a temperature $T_R > m_I$ shortly
after inflation.  If $T_R > \Lambda$, then the effective theory
described in \Eq{eq:exmodel} does not apply, but any renormalizable
ultraviolet completion of these higher dimension operators will
generically induce tree level scattering processes that thermalize
$X$.  On the other hand, if $T_R < \Lambda$, then the higher dimension
operator description is valid, and the interactions in \Eq{eq:exmodel}
will mediate high energy scattering processes such as $ u_i \bar u_j
\rightarrow X_I X_J$, $u_i d_j \rightarrow X_I \bar d_k$, $d_j d_k
\rightarrow X_I \bar u_i$ which also tend to thermalize $X_I$.  The
thermally averaged production cross-section for $X_I$ scales as
$\langle \sigma v\rangle_I \simeq c_I T^2 / \Lambda^4$, where the
proportionality factor $c_I$ depends on $\lambda_{Iijk}$ and
$\kappa_{IJij}$.  Thus, $X$ scattering is dominated by ultraviolet
processes, and is most important at $T_R$.  This effect is familiar
from supersymmetric cosmology, where overproduction of gravitinos
during reheating places a stringent limit on $T_R$
\cite{Moroi:1993mb}.  Similar limits have been computed for a general
hidden sector cosmology \cite{Cheung:2010gj}.

The critical decoupling temperature $T_{D_I}$ defines the temperature
at which these scattering processes go out of equilibrium, {\it
  i.e.}~when $n_{\rm eq} \langle \sigma v \rangle_I \sim H$ where
$n_{\rm eq}$ is the equilibrium number density of $X_I$ and $H$ is the
Hubble parameter.  Together with the scaling of $\langle \sigma v
\rangle_I$, this implies that $T_{D_I}\sim (\Lambda^4/\mPl)^{1/3}$
where $m_{\rm Pl}\simeq 2.4\times 10^{18}~{\rm GeV}$ is the reduced
Planck mass.  In summary, if $T_R > \Lambda$ or $\Lambda > T_R >
T_{D_I}$, then $X_I$ will be thermalized during reheating.  This
thermalization condition is easily satisfied for sufficiently high
values of $T_R$, which we assume for the remainder of our discussion.

While $X_I$ is thermalized initially, it leaves equilibrium once
temperatures drop below $T_{D_I}$.  After decoupling, the yield of
$X_I$ is given is given by
\begin{eqnarray}
Y_I \simeq \frac{n_{\rm eq}(T_{D_I}) }{s(T_{D_I}) },
\label{eq:Y_i}
\end{eqnarray}
where $s$ is the entropy density.  The yield is constant in the
absence of entropy production.  Because we are interested in a case in
which $X_I$ decouples while it is relativistic, we assume throughout
that $T_{D_I} > m_I$.

Once temperatures drop below $m_I$, $X_I$ becomes non-relativistic and
its energy density begins to redshift like matter, since $\rho_I(T)/
s(T)=m_I Y_I$ is a constant.  From this point forward, the energy
density of $X_I$ will evolve to dominate that of any relativistic
species.  During this era, $X_I$ may even come to dominate the total
energy density, at which point the universe will enter a matter
dominated phase.

This period of $X_I$ domination ends when $X_I$ decays via processes
of the form $X_I \rightarrow u_i d_j d_k, \bar{u}_i\bar{d}_j\bar{d}_k,
X_J \bar{u}_i u_j $.  This final state of baryogenesis is similar to
that of \cite{Davoudiasl:2010am}.  The associated partial decay widths
are
\begin{eqnarray}
&&\Gamma(X_I\rightarrow u_i d_j d_k) 
= \frac{|\lambda_{Iijk} |^2}{512\pi^3}\frac{m_{I}^5}{\Lambda^4},
\\
&&\Gamma(X_I\rightarrow X_J \bar{u}_i u_j) 
= \frac{|\kappa_{IJij}| ^2}{1024\pi^3}\frac{m_{I}^5}{\Lambda^4},
\end{eqnarray}
ignoring kinematic factors arising from masses of final state
particles.  The lifetime of $X_I$ is constrained by number of
cosmological constraints.  First, if $\Lambda$ is too high, the model
is constrained by stringent limits from big bang nucleosynthesis (BBN)
\cite{Kawasaki:2004yh,Jedamzik:2006xz} on late time injection of
electromagnetic energy.  The lifetime of $X_I$ is thus bounded by
$\tau_I \lesssim 1 \textrm{ s}$, where
\begin{equation}
\tau_I \simeq 
 \frac{5.2\times 10^{-12}  \textrm{ s}}{  \lambda_I^2}
 \left(\frac{\Lambda}{10^6 \textrm{ GeV}}\right)^4
\left(\frac{1 \textrm{ TeV}}{m_I}\right)^5, 
 \label{eq:longlived}
\end{equation}
and we have defined the effective couplings $\lambda_I^2 = \sum _{ijk}
|\lambda_{Iijk}|^2+\sum _{Jij} |\kappa_{IJij}|^2/4$ where the sums
range over kinematically allowed final states.  \Eq{eq:longlived}
demonstrates that BBN bounds are satisfied for a broad range of
parameter space.  Second, if $\Lambda$ is too low, then $\langle
\sigma v\rangle_I$ will be large and scattering may keep $X_I$ in
equilibrium down to temperatures of order $m_I$.  In this case,
$T_{D_I} < m_I$ and $X_I$ decouples non-relativistically.  While a
residual baryon asymmetry maybe still persist, a correct evaluation
would require a full analysis of Boltzmann equations which goes beyond
the scope of this work, so we restrict to the case where $X_I$
decouples relativistically.

All but the lightest of the $X_I$ will have CP violating decay modes
of different baryon number, so their decays produce a final state
baryon asymmetry through one-loop interference.  The asymmetric width
of $X_I \rightarrow u_i d_j d_k$ is given by interference between tree
and loop diagrams depicted in \Fig{fig:XIudd}.  Ignoring kinematic
factors in the final and intermediate states,
\begin{eqnarray}
\Gamma(X_I\rightarrow u_i d_j d_k) - 
\Gamma(X_I\rightarrow \bar{u}_i \bar{d}_j \bar{d}_k) 
\nonumber \\
=
\sum_{Jl} \frac{{\rm Im}(\lambda_{Iijk}^*\kappa_{IJ li}\lambda_{J l jk})}
{5120\pi^4}
\frac{m_{I}^7}{\Lambda^6},
\end{eqnarray}
where here the sums range over kinematically accessible final and
intermediate states.  We define an asymmetry parameter for each $X_I$
decay by
\begin{eqnarray}
\epsilon_I &=& 
\sum_f B(f) [{\rm BR}(X_I \rightarrow f) - 
{\rm BR}(X_I \rightarrow \bar{f})]
\nonumber \\
&=&\frac{1}{20\pi}
 \frac{\delta_{I}}
 {\lambda_I^2}
 \frac{m_{I}^2}{\Lambda^2},
\label{eq:epsilonB}
\end{eqnarray}
where $f$ sums over final states, ${\rm BR}$ denotes the branching
ratio of a given process, and $B$ is the baryon number of each final
state. Here we have defined the quantity $\delta_I = \sum_{Jijkl} {\rm
  Im}(\lambda_{Iijk}^*\kappa_{IJ li}\lambda_{J l jk})$ to
characterize the net CP violation associated with $X_I$.

In order to compute the net baryon asymmetry, let us first consider
the decay of a single component, $X_I$.  The cosmology depends
sensitively on the relative size of $\rho_I$, the energy density in
$X_I$, and $\rho_R$, the total energy density in radiation, evaluated
just prior to decay.  If $X_I$ decays very soon after it becomes
non-relativistic, then its energy density is of order that of a single
relativistic species, which we dub the ``weak domination regime'',
$\rho_I \ll \rho_R$.  Little entropy is produced and the temperature
of the radiation remains more or less constant.  However, if $X_I$
decays quite late, then it dominates the total energy density of the
universe, which we dub the ``strong domination regime'', $\rho_I \gg
\rho_R$.  Thus, $X_I$ decays will boost the temperature of the
radiation bath to an effective temperature $T_I$ determined by the
total energy density injected, $\rho_I =\pi^2 g_* T_I^4 /30=3H^2
\mPl^2 $ when $\tau_I \sim 1/H$, so
\begin{eqnarray}
T_I \simeq \left(\frac{90}{\pi^2 g_*(T_I)} \right)^{1/4}\sqrt{\mPl/\tau_I},
\label{eq:TX}
\end{eqnarray}
where $g_*$ counts relativistic degrees of freedom.

The asymmetric baryon number generated by $X_I$ decays is given by
\begin{eqnarray}
\label{eq:etaB}
\eta_I &=& \epsilon_I Y_I d_I,
\end{eqnarray}
where $Y_I$ is defined in \Eq{eq:Y_i} and the dilution factor $d_I$ is
the ratio of the entropy density before and after $X_I$ decays, so
\begin{eqnarray} 
d_I &\simeq& \left\{ \begin{array}{ll}
    1&,\quad  \rho_I \ll \rho_R \\
    \frac{3T_I}{4m_I Y_I}&,\quad \rho_I \gg \rho_R \\
\end{array}\right. ,
\label{eq:dilution}
\end{eqnarray} 
so the dilution factor is much smaller in the strong domination regime.
Applying Eqs.~(\ref{eq:TX}), (\ref{eq:etaB}) and (\ref{eq:dilution})
we find
\begin{eqnarray}
\eta_I \simeq 
\frac{6.2 \times 10^{-11} }{\lambda_I^2/\delta_I}
\left(\frac{10^3 \textrm{ TeV}}{\Lambda}\right)^2
\left(\frac{m_I}{1 \textrm{ TeV}}\right)^{2} 
\left(\frac{106.75}{g_*(T_{D_I})}\right), \nonumber \\
\label{eq:etaBobs2}
\end{eqnarray}
in the weak domination regime, $\rho_I \ll \rho_R$.   Meanwhile,
\begin{eqnarray}
\eta_I \simeq 
\frac{3.5\times 10^{-9} }{  \lambda_I/\delta_I}
\left(\frac{10^3 \textrm{ TeV}}{\Lambda}\right)^4
\left(\frac{m_I}{1 \textrm{ TeV}}\right)^{7/2} 
\left(\frac{106.75}{g_*(T_{I})}\right)^{1/4},\nonumber \\ 
\label{eq:etaBobs}
\end{eqnarray}
in the strong domination regime, $\rho_I \gg \rho_R$.  Note that
sphaleron processes will partially wash out the baryon asymmetry if
$T_{I}\gtrsim m_W/\alpha_W$, where $m_W$ is the $W$ boson mass,
$\alpha_W=g^2/4 \pi$, and $g$ is the SU(2)$_L$ gauge coupling.  In
this case, the net baryon asymmetry is processed according to $\eta_I
\rightarrow (28/79)\eta_I$.

With the expressions in Eqs.~(\ref{eq:etaBobs2}) and
(\ref{eq:etaBobs}), it is straightforward to compute the net baryon
asymmetry generated from all $X_I$ decays.  If the masses and
couplings are not hierarchical, then each $X_I$ should decay around a
similar time.  In this case, the $X_I$ should either all be in the
weak domination regime or all be in the strong domination regime.  For
the former, little entropy is produced by each $X_I$ decay, and the
net baryon asymmetry is simply given by the sum of all $\eta_I$.  For
the latter, entropy is substantially produced in each decay, thus
diluting the asymmetry generated in earlier epochs.  In this case the
net baryon asymmetry is dominated by $\eta_I$ from latest of the $X_I$
decays.

Finally, let consider the issue of DM.  Let us denote a stable
component of the $X_I$ multiplet by $X_{\rm DM}$.  If $X_{\rm DM}$ is
lighter than the proton, then $X_{\rm DM}$ is exactly stable because
it is the lightest odd particle under the unbroken $\mathbb{Z}_2$
subgroup of baryon number.  The primordial relic abundance of $X_{\rm
  DM}$ is $\Omega_{\rm DM} = m_{\rm DM} Y^{\rm tot}_{\rm
  DM}(s/\rho_c)_0$, where $(\rho_c/s)_0\simeq 3.6~h^2\times 10^{-9}
~{\rm GeV}$ for $h\simeq0.67$ \cite{Planck}.  The DM abundance arises
from two sources, $Y^{\rm tot}_{\rm DM} = Y_{\rm DM}^{\rm th} + Y_{\rm
  DM}^{\rm dec}$ arising from thermal scattering during initial
reheating and decays of heavier $X_I$, respectively.  Assuming that
baryogenesis is dominated by the decays of a single species $X_I$,
these contributions are
\begin{eqnarray}
&&Y_{\rm DM}^{\rm th} 
= \frac{\eta_I}{\epsilon_I} \frac{Y_{\rm DM}} {Y_I} ,
\\ &&
Y_{\rm DM}^{\rm dec}= \frac{\eta_I}{\epsilon_I} 
{\rm BR}({X_I\rightarrow X_{\rm DM}}),
\end{eqnarray}
where $Y_{\rm DM}$ and $Y_I$ are as defined in \Eq{eq:Y_i}. 
Typically, $Y_{\rm DM}$ and $Y_I$ will be comparable, and BR$(X_I
\rightarrow X_{\rm DM}) \sim {\cal O}(1)$, so the
contributions to the DM abundance from thermal scattering and decays
will be of similar order, but both $\sim 1/ \epsilon_I$ larger than
the asymmetric yield.

\begin{figure}[t]
  \begin{center}\hspace*{-.75cm}
    \includegraphics[scale=0.55]{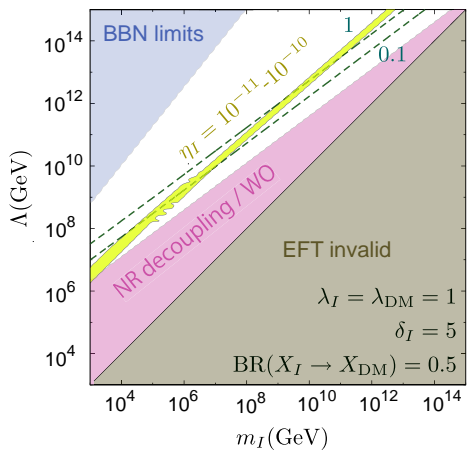}
  \end{center}
  \vspace*{-.5cm}
  \caption{Cosmologically allowed regions of parameter space.  The
    grey region lies outside the regime of the effective theory.  The
    blue region is disfavored by BBN, while the purple region is
    excluded by the requirements that $X_I$ decouple relativistically
    and washout be evaded.  The yellow region accommodates the
    observed primordial baryon asymmetry, while the green dashed lines
    denote the required DM mass in keV.  }
  \label{fig:eta}
\end{figure}

Finally, we summarize the allowed parameter space for a single species
of $X_I$ in \Fig{fig:eta}.  The grey region indicates where the higher
dimension operator description is invalid because $m_I > \Lambda$. For
$\lambda_I =1$, the blue region depicts the parameter space excluded
by BBN limits on late decays of $X_I$, while the purple region depicts
the parameter space excluded by requiring that $X_I$ decouples
relativistically---{\it i.e.}~it is not thermalized by scattering
processes at temperatures of order $m_I$.  The purple region is very
similar to the region excluded by washout from scattering processes,
assuming $c_I = \lambda_{I} / 4\pi$ for all $X_I$.  Furthermore,
taking that $\delta_I= 5$ and that $X_I$ is the primary origin of
baryogenesis, then the yellow band indicates where $10^{-11}\le
\eta_I\le 10^{-10}$.  Note that this choice for $\delta_I$ is not in a
strong coupling regime because all couplings are really normalized to
a higher dimension operator scale $\Lambda$.  As noted earlier, the
model also has the option of including primordial relic DM.  Requiring
that $\Omega_{\rm DM}h^2\simeq 0.11$~\cite{Planck} fixes the DM mass,
which is denote by green dashed lines for $m_{\rm DM} = 0.1$ and
$1~{\rm keV}$.  Thus, for ${\cal O}(1)$ couplings, the observed baryon
asymmetry is generated in the regime in which our effective theory
analysis is valid, and DM can also be accommodated.

\begin{center}
\textbf{\textsc{III. Experimental Signatures}} 
\end{center}

Our proposal offers experimentally observable consequences connected
with the operators directly involved in asymmetry generation.  Baryon
number violating operators will typically induce highly constrained
$n$-$\bar n$ oscillations via the effective operator ${(u
  d)(dd)(ud)}/{M_{n\textrm{-}\bar{n}}^5}$, where $u$ and $d$ are the
right-handed up and down quarks, while Lorentz indices are contracted
within the parentheses and color indices are contracted in the unique
way.  This operator is not induced at tree level in the model defined
in \Eq{eq:exmodel}, due to an accidental antisymmetry in the flavor
indices of the coupling constant $\lambda_{Iijk}$.  However, this
operator will be induced at loop order, and more generally will be
present at tree level if there are higher dimension operators in
addition to those in \Eq{eq:exmodel}.  For example $n$-$\bar n$
oscillation will be induced if there are operators of the form
$\lambda'_{Iijk}(X_I d_j) (d_k u_i)/\Lambda^2$.  Integrating out $X_I$
will produce the $n$-$\bar n$ operator with an effective cutoff
$M^5_{n\textrm{-}\bar{n}} \sim { \Lambda^4 m_I}/{\lambda'^2_{I111} }$.
The characteristic time scale of $n$-$\bar n$ oscillation goes as
$\tau_{n\textrm{-}\bar n} \sim M^5_{n\textrm{-}\bar{n}} /(3\times
10^{-4} \textrm{ GeV}^6)$ \cite{Goity:1994dq,Csaki:2011ge}, which
together with the experimental bound, $\tau_{n\mathchar`-\bar{n}}\ge
2.4\times 10^8~{\rm s}$~\cite{Abe:2011ky} implies
\begin{eqnarray} 
\Lambda \gtrsim
3.2\times 10^6 \textrm{ GeV} \;  {|\lambda'_{I111}|}^{1/2} \left(\frac{1
    \textrm{ TeV}}{m_I}\right)^{1/4}, 
\end{eqnarray} 
so $n$-$\bar n$ oscillations could offer a sensitive probe of the low
scale variants of this baryogenesis mechanism.

Flavor violation offers another possible probe of this model.  In
particular, $K^0\mathchar`-\bar{K}^0$ mixing is mediated by the
operator $(dd)(\bar s\bar s)/M_{K^0\textrm{-}\bar K^0}^2$, which is
induced at loop-level, where $M_{K^0\textrm{-}\bar K^0}^2 \sim 16
\pi^2 \Lambda^4 / \lambda'^*_{I111}\lambda'_{I122}m_I^2$. Comparing
the estimated mixing rate with the experimental bound, ${\rm Im}\,
M_{12}\le
3.3\times 10^{-18}~{\rm
  GeV}$~\cite{PDG,Buras:2010pza,Laiho:2009eu,Mescia:2012fg}, gives
\begin{eqnarray}
\Lambda \gtrsim 4.4\times 10^4~{\rm GeV}\;
{\rm Im}(\lambda'^*_{I111}\lambda'_{I122})^{1/4}
\left(\frac{m_I}{1~{\rm TeV}}\right)^{1/2},
\end{eqnarray}
which can be competitive with $n$-$\bar n$ limits.

In theories where there exists a cosmologically stable dark matter
candidate $X_{\rm DM}$, there are stringent limits on proton decay via
the process $p\rightarrow \pi^+ X_{\rm DM}$, whose decay rate is
estimated as $\Gamma(p\rightarrow \pi^+ X_{\rm DM}) \sim \lambda_{\rm
  DM}^2 m_p{\Lambda}^4_{\rm QCD} /16\pi \Lambda^4$, where $m_p$ is the
proton mass and ${\Lambda}_{\rm QCD}\sim 250~{\rm MeV}$ is the QCD
scale~\cite{Aoki:2008ku}. Then the experimental bound,
$\tau_{p\rightarrow \pi^+ \nu}\ge 2.5\times 10^{31}~{\rm
  yr}$~\cite{PDG}, gives a very stringent limit on the cutoff
\begin{eqnarray}
\Lambda\gtrsim
5.5 \times 10^{14}~{\rm  GeV} \;
{\lambda_{\rm DM}^{1/2}}
\left(\frac{{\Lambda}_{\rm QCD}}{250~{\rm MeV}}\right).
\end{eqnarray}
In order to evade the proton decay bound for the DM model, we must
assume a hierarchical flavor structure in the coupling $X_{\rm DM}$ to
the light quarks.  This can be accommodated in
models of minimal flavor violation (MFV), {\it e.g.}~in R-parity
violating supersymmetric theories \cite{Nikolidakis:2007fc,
  Csaki:2011ge}.

\begin{center}
\textbf{\textsc{IV. Conclusions}} 
\end{center}

In this letter we propose a simple baryogenesis model comprised of the
standard model coupled via higher dimension operators to a neutral
multiplet $X$.  In the early universe, $X$ will typically be
thermalized after reheating, but it eventually decouples and becomes
non-relativistic.  Subsequently, $X$ evolves into a sizable fraction
of the total energy density of the universe until it decays out of
equilibrium, yielding a baryon asymmetry and possibly dark matter.
This scenario can be experimentally probed via $n$-$\bar n$
oscillations, flavor violation, and proton decay.

\vspace*{0.05cm}

\noindent {\it Acknowledgments}: This publication is funded by the
Gordon and Betty Moore Foundation through Grant GBMF \#776 to the
Caltech Moore Center for Theoretical Cosmology and Physics. The work
is also supported by the U.S. Department of Energy under Contract
No. DE-FG02-92ER40701.

\appendix

\end{document}